\documentclass[runningheads]{llncs}
\usepackage{graphicx}
\usepackage[table]{xcolor}
\usepackage{stackengine}
\usepackage{amsmath,amssymb}
\usepackage{hyperref}
\usepackage{flushend}
\usepackage{comment}

\usepackage[utf8]{inputenc}
\usepackage{fourier, erewhon}
\usepackage{geometry}
\usepackage{array, caption, floatrow, tabularx, makecell, booktabs}%
\setcellgapes{3pt}

\usepackage{graphics} 
\usepackage{mathptmx} 
\usepackage{times} 
\usepackage{amsmath} 
\usepackage{amssymb}  
\usepackage[utf8]{inputenc}
\usepackage[english]{babel}
\usepackage[export]{adjustbox}

\usepackage{fancyhdr}
\title{Patch Clustering for Representation of Histopathology Images}

\author{Wafa Chenni$^1$, Habib Herbi$^2$, Morteza Babaie$^3$, H.R.Tizhoosh$^{3,4}$}
\institute{$^1$ Pierre and Marie Curie University, Paris, France\\
$^2$ Sorbonne University, Paris, Ile-de-France\\
$^3$ Kimia Lab, University of Waterloo, Canada\\
$^4$ Vector Institute, Toronto, Canada}

\begin{document}

\maketitle

\begin{abstract}
Whole Slide Imaging (WSI) has become an important topic during the last decade. Even though significant progress in both medical image processing and computational resources has been achieved, there are still problems in WSI that need to be solved. A major challenge is the scan size. The dimensions of digitized tissue samples may
exceed 100,000 by 100,000 pixels causing memory and efficiency obstacles
 for real-time processing. The main contribution of this work is representing a WSI by selecting a small number of patches for algorithmic processing (e.g., indexing and search). As a result, we reduced the search time and storage by various factors between ($50\% - 90\%$), while losing only a few percentages in the patch retrieval accuracy. A self-organizing map (SOM) has been applied on local binary patterns (LBP) and deep features of the KimiaPath24 dataset in order to  cluster patches that share the same characteristics. We used a Gaussian mixture model (GMM) to represent each class with a rather small (10\%-50\%) portion of patches. The results showed that LBP features can outperform deep features. By selecting only 50\% of all patches after SOM clustering and GMM patch selection, we received 65\%  accuracy for  retrieval of the best match, while the maximum accuracy (using all patches) was 69\%.
\end{abstract}

\section{Introduction}

The advances in digital image processing and machine learning for digital pathology are showing practical results. The advantage of such techniques is the ability to assist pathologists for higher accuracy and efficiency. Such algorithms lead to more reliable diagnosis by presenting computer-based second opinions to the clinician \cite{robboy2016retrieval}. Digital Pathology (DP) uses Whole Slide Imaging (WSI) as a base for diagnosis. Unlike the traditional pathology workflow in which the tissue samples are inspected under a microscope and stored in  physical archives, WSI enables the digitization of glass slides to very high-resolution digital images (slides/scans). The introduction of such technologies has led to the development of countless methods combining machine learning and image processing to support the  diagnostic workflow which is labour-intensive, time costly, and subject to human errors \cite{alzubaidi2017computer}. The digitization of the biopsy samples has simplified parts of the analysis, however, it has also introduced several challenges. There are only a few public digital datasets available for machine-learning purposes  \cite{tizhoosh2018representing}. In addition, the existing datasets are generally unlabeled because of the tedious and costly nature of the manual delineation of regions of interest in digital images. Moreover, DP methods suffer from the image  imperfections caused by the presence of artifacts and the absence of accurate methods for tissue (foreground) extraction \cite{moriya2018unsupervised}.
Content-based image retrieval (CBIR) is considered as a practical solution for  processing unlabeled data. Retrieving similar cases from pathology archives alongside their treatment records may help pathologists to write their reports much more confidently.  
Finally, the requirements of WSI for memory usage and computational power is problematic for IT infrastructures of hospitals and clinics. Therefore, it is desired to have  solutions that make the image processing more memory efficient and computationally less expensive.
This paper addresses the reduction of data dimensionality by clustering images with in order to provide a compact representation of the scans for algorithmic processing. Our techniques are developed under the constraint of working with unlabeled data, a constrained that is motivated by the reality of the clinical workflow. 

\section{Related Works}
Tissue examination under a microscope reveals important information to render accurate diagnosis and thus, provide effective treatment for different diseases \cite{da0}. DP offers several opportunities and also presents challenges to the image processing architectures \cite{ref71}. Presently, only a small fraction of glass slides are digitized \cite{da1}, but even if WSI was more widely available, there are a number of technical issues that would need to be addressed for their effective usage. One of the main challenges is data management and storage \cite{da0}. Most importantly, the large dimensions of the WSI files require a large amount of memory and a expensive computational power.

Content-Based Image Retrieval (CBIR) is an approach to find images with similar visual content to a query image by searching a large archive of images. This is helpful in medical imaging and DP databases where text annotations alone might be insufficient to precisely describe an image \cite{ref105,ref107}. In order to retrieve similar images, a proper feature representation is needed \cite{khatami2018parallel}. 
 In CBIR, accuracy and fast search for similar images from large datasets are important. Therefore, various techniques for dimensionality reduction of features are used to speed up CBIR systems \cite{2017retrieving}. Some of these techniques  include principal component analysis (PCA), compact bilinear pooling \cite{ref115} and fast approximate nearest neighbor search \cite{ref116}. 
Image subsetting methods \cite{1658090,Aiad2009} have been used to choose a small region of the whole slide images for computational analysis while reducing the size of the image for a better tissue representation. Other image subsetting algorithms use sparsity models for multi-channel representation and classification, and expectation maximization by logistic regression \cite{srinivas2014simultaneous,hou2016patch}. Generally, the $20x$ magnification is commonly used for many diagnostic tasks \cite{ref15}, \cite{ref16}. As well, dividing the whole slides into small patches (or tiles) of 256$\times$256 to 1000$\times$1000 pixels is a common strategy to overcome the large dimensionality of the WSI data \cite{ref61,ref63}. This approach results in thousands of patches that should be analyzed individually. Low-resolution approaches are considerably faster, however, they may loose the local morphology. One possible solution is regional averaging where a region is not considered region of interest (ROI) unless it extends over multiples patches. On the other hand, this can cause missing small ROIs such as small or isolated tumors. Another solution would be to analyze the complete image on low resolution and then refine this result on high-resolution patches by using a  registration on each patch \cite{ref49}. In this manner, the local morphology is taken into account. However, one major downside is the significantly longer run time. In this work, We propose unsupervised learning using handcrafted and deep features, followed by patch selection through Gaussian mixture models, to provide a more compact representation of the digital slides for image indexing and search purposes. 

\section{Materials and Methods}

\begin{figure}[tb]
    \centering  {
\stackunder[5pt]{\includegraphics[width=0.22\columnwidth,height=0.22\columnwidth]{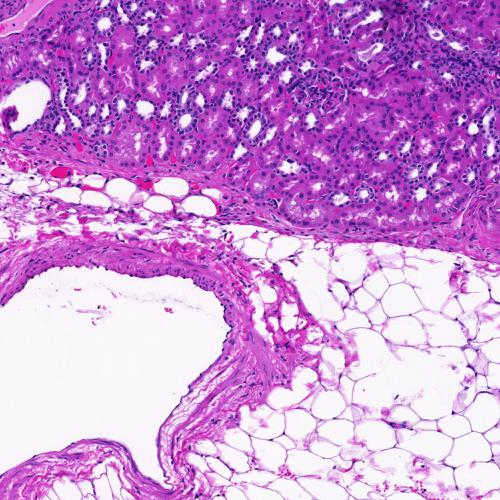}}{\tiny }
\stackunder[5pt]{\includegraphics[width=0.22\columnwidth,height=0.22\columnwidth]{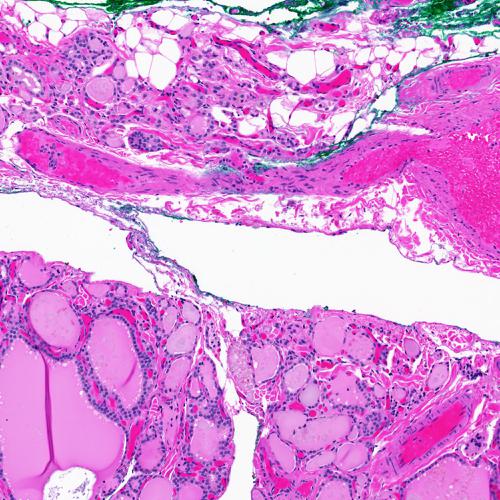}}{\tiny }
\stackunder[5pt]{\includegraphics[width=0.22\columnwidth,height=0.22\columnwidth]{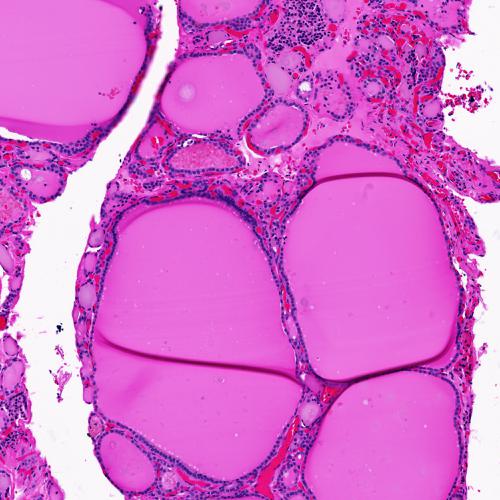}}{\tiny }
\stackunder[5pt]{\includegraphics[width=0.22\columnwidth,height=0.22\columnwidth]{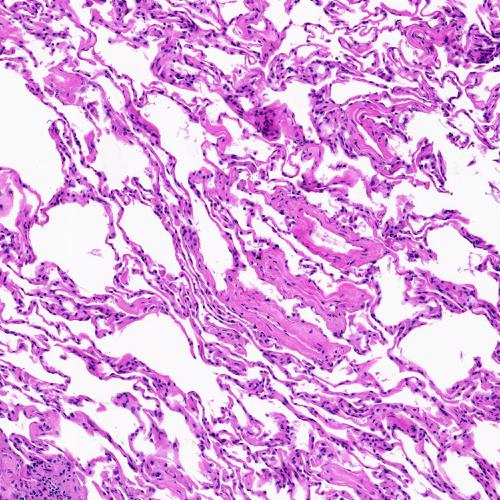}}{\tiny }
\label{fig:samples1}}
 \protect\caption{Sample patches from KimiaPath24 dataset.}
  \label{fig:samples}
\end{figure}

\textbf{Dataset and Data Preparation --} We used the KimiaPath24 dataset to evaluate our experiments. This dataset contain 24 WSIs. The slides show diverse organs and tissue types with different texture patterns \cite{babaie2017classification}. The glass slides were captured by a digital scanner in bright field using a 0.75 NA lens \footnote{TissueScope LE scanner by Huron Digital Pathology}. The dataset contains 1325 test images (patches) of size 1000 $\times$ 1000 pixels (0.5mm $\times$ 0.5mm) from all 24 cases. Fig. \ref{fig:samples} shows some example patches (the dataset can be downloaded online\footnote{http://kimia.uwaterloo.ca/kimia\_lab\_data\_Path24.html}). 

All training and test patches are down-sampled from $1000 \times 1000$ pixels to $250 \times 250$ pixels in order to be more easily processed by for the feature extraction. We patched WSIs without overlap and then we removed all patches with high background homogeneity (more than $\%99$)  \cite{babaie2017classification}. As a result, we created  27,055 training patches from 24 WSIs. The presented dataset comprising of diverse body parts may be suitable for intra-class search operations such as metastasis and floater detection.  

\begin{figure*}[tb]
    \centering    
        \includegraphics[width=0.9\textwidth]{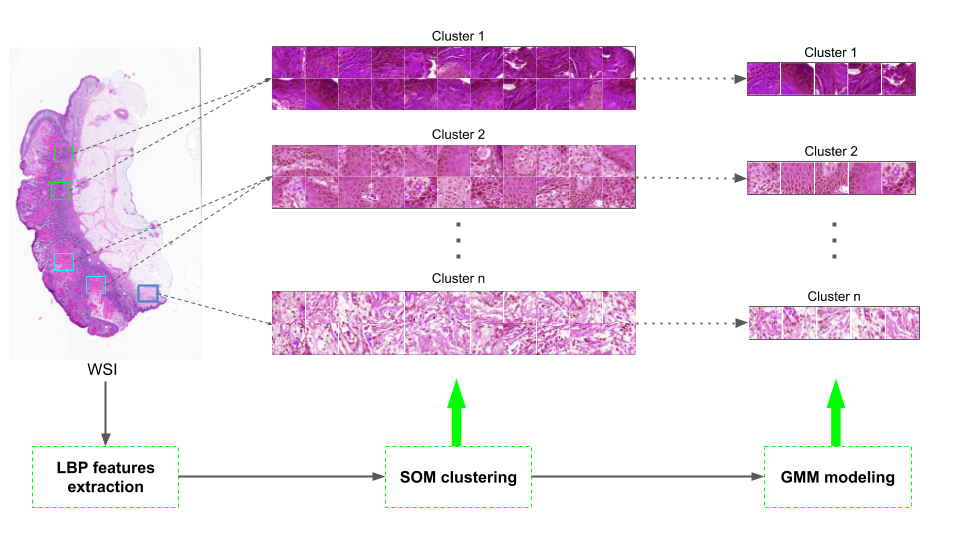}
\caption{SOM clustering and GMM patch selection.}
    \label{fig:samplesrbgrgf}
\end{figure*}

\textbf{Methodology --}  Fig. \ref{fig:samplesrbgrgf} illustrates our approach. We divided the whole slide image into many patches, extract features, cluster the patches and then selected a subset of patches to represent the scan. We applied image search to verify the accuracy loss as a consequence of data reduction. We have performed search by extracting same features from the test set and then compared them against features from all training cases by calculating the Euclidean distance as a measure of (dis)similarity. The most similar patch is considered to be the output of the CBIR system. 

Convolutional Neural Networks (CNN) can learn general features that are not specific to the dataset or task \cite{cnn}. The deeper layers are more specific to the task of the network. Many results indicate that extracted features from CNNs (deep features) are highly discriminative \cite{ref433}. These features are extracted from different layers of the CNN depending on the degree of specificity of the feature \cite{ref433_2}. Usually, the features are extracted from the last layer before the classification layer which allows getting the most specific and high abstraction features that can be used for another task (dimensionality reduction, unsupervised learning, etc.).
In this work, we used all 4096 outputs of the last layer before the classification layer in the VGG16 network, a  pre-trained network model with 16 layers  \cite{han2015deep,ref433_2}. The second feature extraction method is a handcrafted method that uses the LBP algorithm (local binary patterns) \cite{lbp0,lbp1}. The obtained feature vectors are histograms of uniform and rotation-invariant patterns. The LBP vector is a concatenation of two vectors. The first one has a radius  parameter of 3 pixels and 24 pixels to consider resulting in 26 bins. The second one has a radius  parameter of 1 pixel and 8 pixels to consider set to 8 resulting in 10 bins. The  concatenated  histogram will be of 36 dimensions (bins) for  each  patch.

\begin{figure}[tb]
\centering
\stackunder[5pt]{\includegraphics[width=0.33\columnwidth,height=0.33\columnwidth]{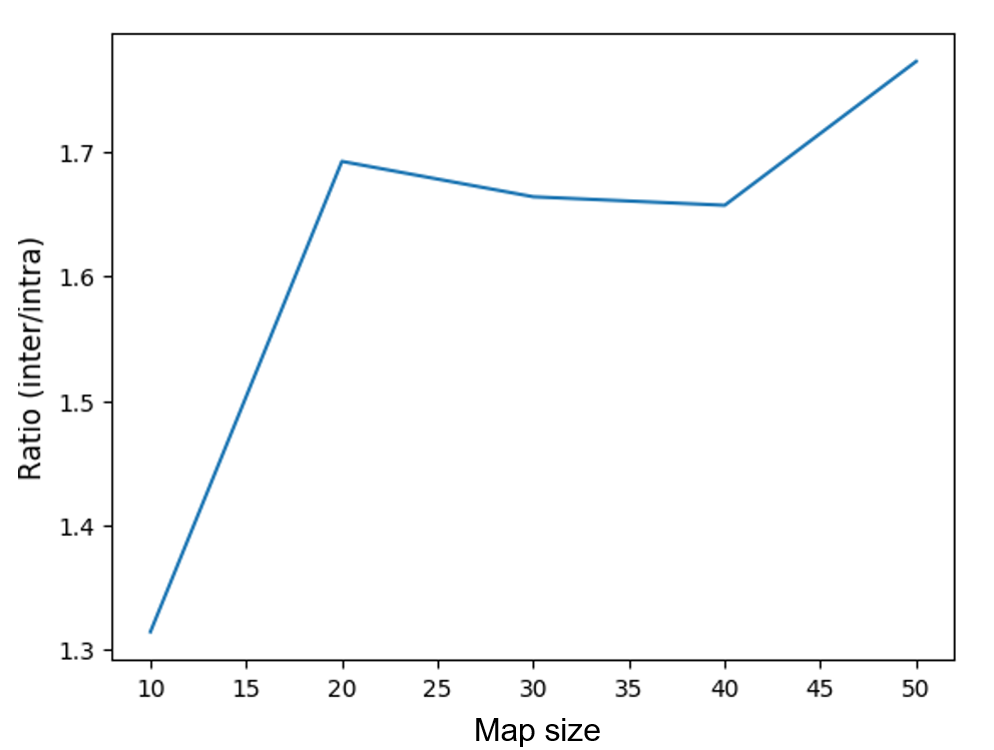}}{\tiny }
\stackunder[5pt]{\includegraphics[width=0.33\columnwidth,height=0.33\columnwidth]{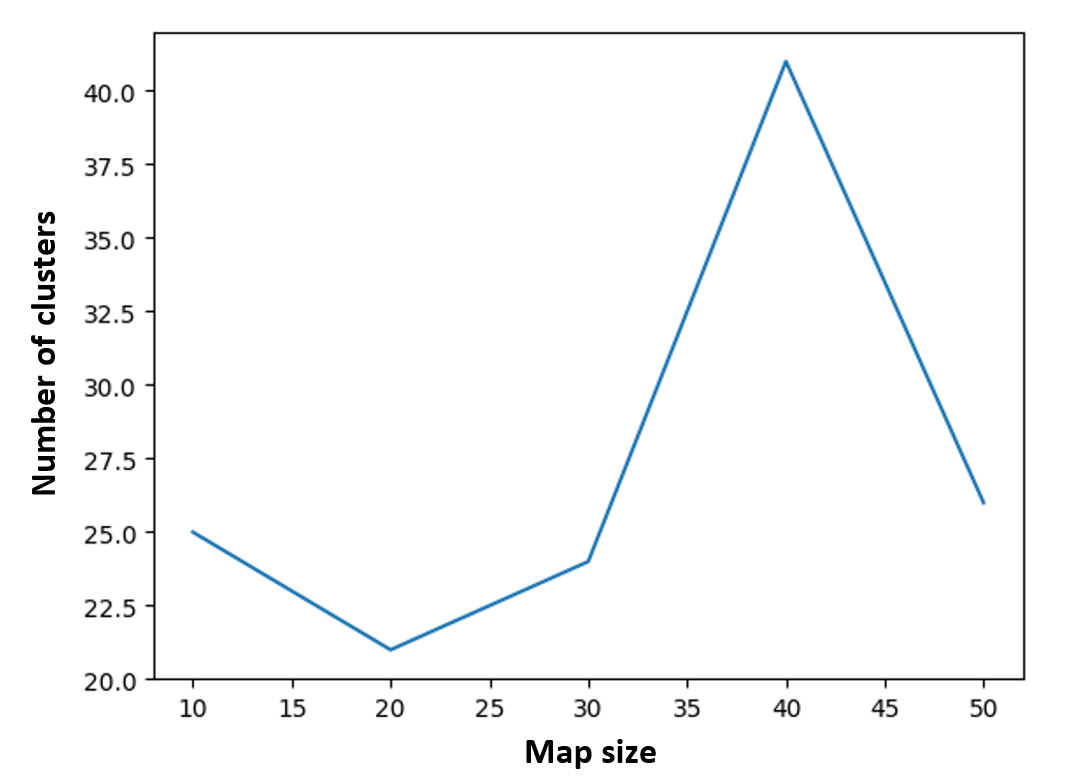}}{\tiny }

\caption{SOM parameters: The effect of map size on ratio and number of clusters; Left: variance ratio versus map size. Right: the number of clusters versus the map size. }
\label{fig:som_param1}
\end{figure}

All patches of a scan are represented by two different sets of features for comparison, namely deep features and LBP histograms. We then train SOM to cluster each patch. We do not know how many clusters each scan may contain. Hence, each scan is split into a given number of clusters found by the SOM algorithm. The range of number of clusters found by SOM was between 10 and 20. Parameter tuning is performed to shed light on variance and map size (see Fig. \ref{fig:som_param1}). 
We used GMMs \cite{ref72} for patch selection. The number of representatives for each cluster is investigated from range $10\%$ to $50\%$ of the total number of the patches. It is important to point out that the deep features have a large feature vector (more than 4000 elements). Therefore, we used PCA to reduce the dimensionality of deep features. We kept 95$\%$ of the variance for each vector which yielded a new feature vector of 1078 elements. We also experimented with random patch selection which provided slightly worse results compared to GMMs.

While trying to minimize the number of clusters and maximize the variance ratio, we observed that these two parameters are positively correlated. Fig. \ref{fig:som_param1} shows the variation of these two parameters versus the map size. We can see that with an increasing ratio, the number of clusters is reaching large numbers. Based on empirical knowledge, a  desirable number of clusters would be less than 30. We can see that 20 may be regarded as a suitable value for the map size as it provides a good compromise between the number of clusters and the inter-/intra-variance ratio. 
It is important to point out that changing the SOM's learning rate did not have much impact. With these values, we are able to cluster 18 clusters per scan on average. Some of the clusters contain few patches (less than 1$\%$ of the total number of patches). We  merged such clusters with the closest cluster (using Euclidean distance) resulting in a smaller  number of clusters. In case of important clusters removed by merging, GMM may still select those patches. However, one must keep in mind that main purpose of the CBIR systems is generally recognizing dominant tissue patterns and not detecting minute cellular details. The latter is a subject for detection and segmentation algorithms. 


\section{Results}
\begin{figure*}[htb]
    \centering    {
        \includegraphics[width=0.6\textwidth]{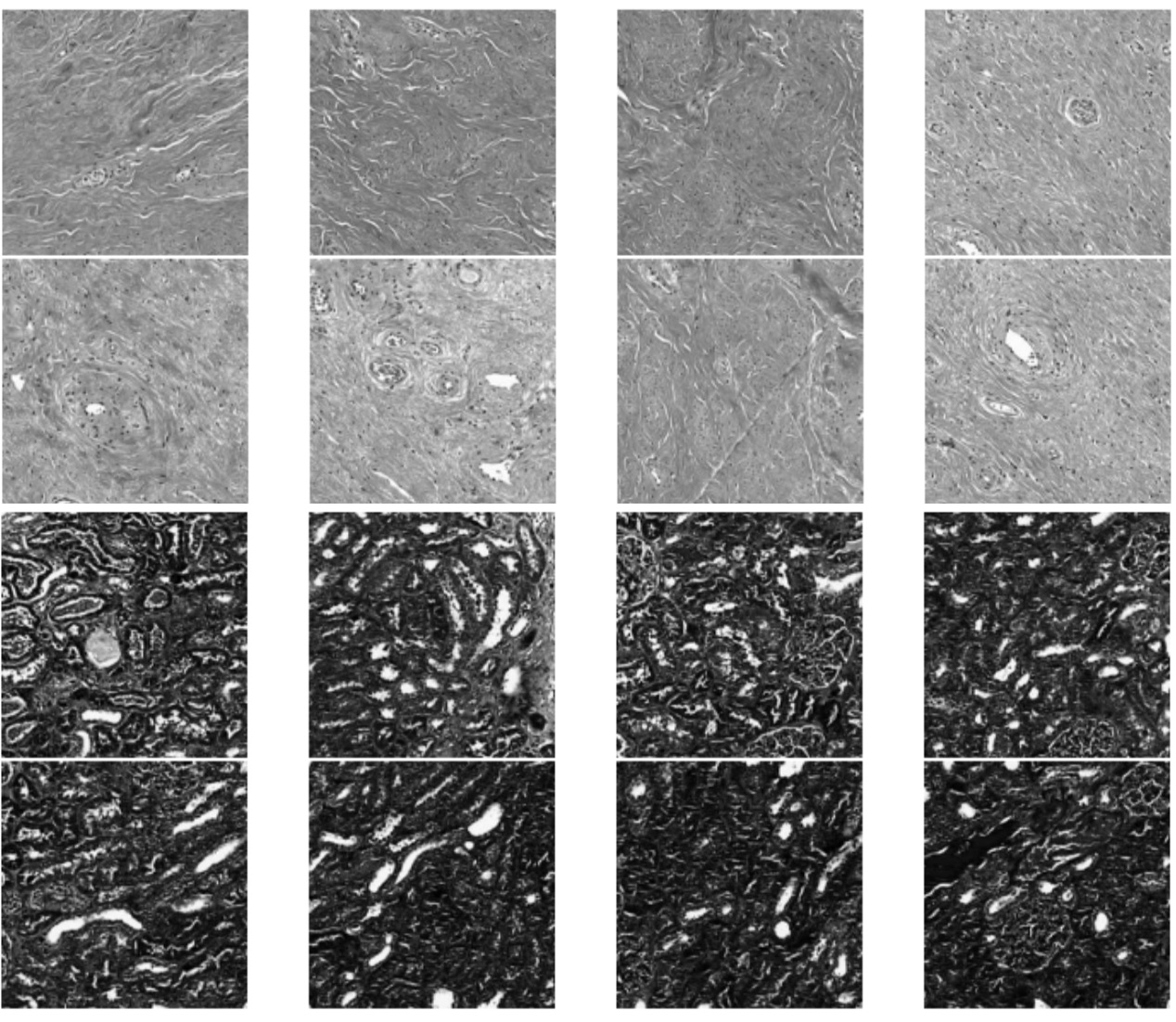}
        \label{fig:ergrsamples1}    }
 \protect\caption{Two sample SOM clusters  using deep features.}
    \label{fig:somdeep}
\end{figure*}

\begin{figure*}[htb]
    \centering    {
        \includegraphics[width=0.6\textwidth]{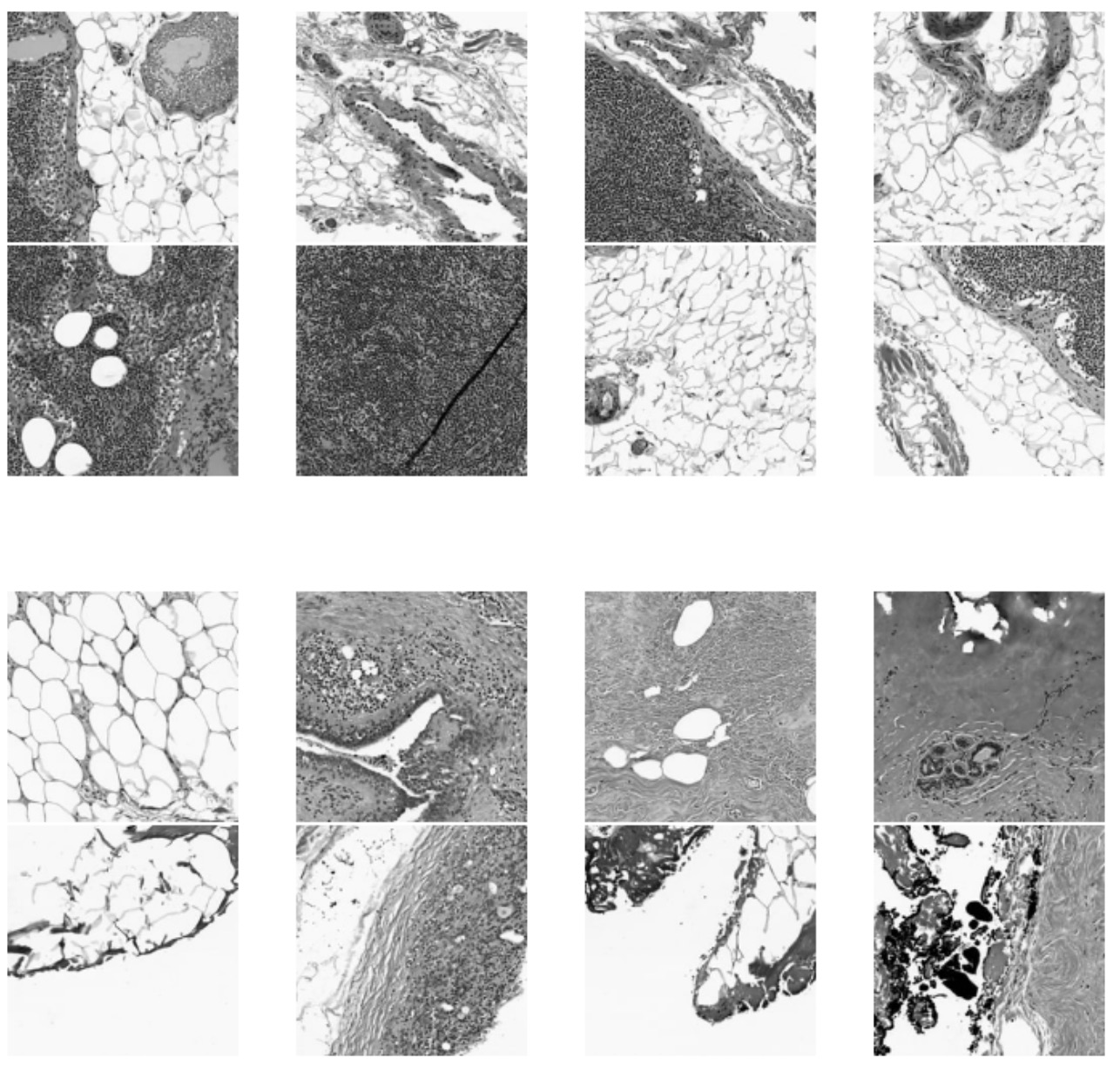}
        \label{fig:ergrsamples1}    }
 \protect\caption{Two sample GMM patch selection.}
    \label{fig:gmmgroups}
\end{figure*}

\begin{figure}[htb]
\includegraphics[width=0.65\textwidth,center]{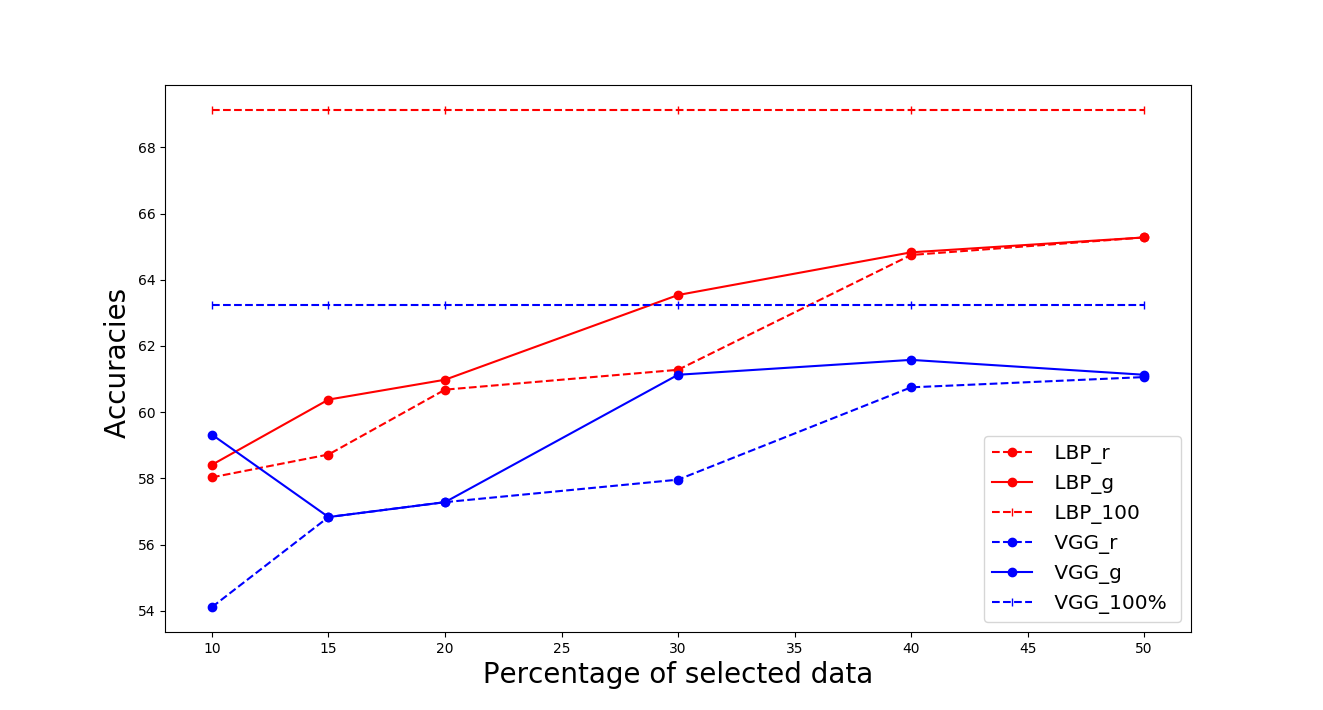}
\caption{Evolution of performance of our methods versus percentages of reduced data. Note that the underscore '100$\%$' means that the entire training data set has been used.}
\label{fig:plot_accuracies}
\end{figure}

LBP descriptor outperformed deep features 3 times out of 4. Other (deeper) networks may perform better, however, they also require more resources. The LBP histogram has 36 bins while the VGG16 feature vector length is 1078 (after application of PCA). Fig. \ref{fig:somdeep} shows two examples for sample patches that SOM groups together using deep features. Fig. \ref{fig:gmmgroups} illustrates two sets of patches selected by GMMs from SOM clusters. The accuracy calculation aims to compare the performance of the proposed method for image retrieval using LBP and deep features separately by comparing it to the accuracy obtained using the  training data set (27,055 patches). We have used the KimiaPath24 guidelines to calculate the  \textbf{patch-to-scan accuracy} $\eta_\textrm{p}$, \textbf{whole-scan accuracy} $\eta_\textrm{W}$ and the total accuracy $\eta_\textrm{total}$. LBP's performance improved with the increase of selected data whereas for VGG16 features the performance only improved with the increase of selected data in the case of random selection. Indeed, with GMM selection, performance decreased with more data. This might be due to the length of VGG feature vectors. Fig. \ref{fig:plot_accuracies} gives a general overview of the performances evaluation while Table \ref{tab:VGG_distances} reports all accuracy measurements (\_r and \_g are indicate random selection and GMM selection, respectively).

\begin{table}[htb!]
 \begin{tabular}{c|c|c|c|c|c}
     \hline  $\quad$GMM Selection$\quad$ & $\quad$ Features $\quad$   & $\quad$Feature Length$\quad$ &   $\quad \eta_\textrm{p}\quad$    &  $\quad\eta_\textrm{W}\quad$   & $\quad\eta_\textrm{total}\quad$   \\ 
        \hline   10\% &  LBP & 36 & 58.41 & 58.03 & 33.7  \\ 
  10\% & VGG  & 1078  & 59.32 & 61.47 & 36.46   \\ 
    
   \hline  15\% &  LBP & 36  & 60.38 & 60.87 & 36.75  \\  
  15\% &  VGG & 1078  &  57.28 &  57.91 & 33.17  \\ 
  
   \hline   20\% &   LBP  & 36 & 60.98 & 61.82 & 37.7  \\ 
 20\% &  VGG  & 1078 & 57.28 & 59.51 & 34.08   \\ 
    \hline 30\% &   LBP & 36  & 63.54 & 63.27 &  40.21  \\ 
  30\% &   VGG & 1078  & 57.96 & 58.72 &  34.03  \\
    \hline   40\% &  LBP & 36  & 64.83 & 64.98 &  42.13 \\
   40\% & VGG  & 1078 & 61.58 & 64.01 &  39.42\\
    
      \hline  50\% & LBP & 36  & 65.28 & 64.30 &  41.98 \\
     50\% &  VGG  & 1078  & 61.13 & 63.33 &  38.71 \\
      
    \hline 100\% &  LBP & 36 & 69.13 & 69.40 & 47.98  \\ 
  100\% &   VGG  & 1078  & 63.25 &  66.19 & 41.86 \\  
   100\%  &  LBP  \cite{babaie2017classification} & 555  & 66.11  
 &  62.52 & 41.33 \\    
     \hline
      \end{tabular}
\caption{Retrieval accuracy for VGG16 and LBP features with the GMM patch selection.}
      \label{tab:VGG_distances}

\end{table}%

\section{Conclusions}
Performance of both LBP and deep features generally drops as a result of patch selection, a fact that can be considered during the algorithm design. However, the run time and memory requirements can be considerably reduced  which can be an  advantage in dealing with large WSI archives. For CBIR systems in histopathology, retrieval of similar images is a major challenge because of the enormous size of the archives. The results of our experiments showed that for the algorithmic purposes such as image search the size of the image indexing (i.e., feature calculation) can be drastically reduced while keeping the relevant information and characteristics of each scan. Keeping\textbf{ 50\%} of the patches and using LBP descriptor and GMM selection reduces the index size and, expectedly, the computational requirements by 50\% and reaches a CBIR accuracy of \textbf{65\%} (for the first match) only \textbf{4\%} less than feature extraction for the entire data.

\bibliographystyle{splncs04}
\bibliography{mybibliography}

\end{document}